\documentclass[onecolumn]{emulateapj}

\usepackage{graphicx,natbib}
\usepackage{hyperref}
 \shorttitle{TIC 452991707 \& TIC 452991693: system with three eclipsing binaries}
 \shortauthors{Zasche \& Henzl}

\begin{document}

%\begin{Titlepage}

\title{\bf{\large TIC 452991707 \& TIC 452991693 as a candidate \\[2mm] sextuple system with three eclipsing
binaries}}

\author{P. Zasche\altaffilmark{1}
          \and
        Z.Henzl\altaffilmark{2} }

 \affil{
  \altaffilmark{1} Charles University, Faculty of Mathematics and Physics, Astronomical Institute,
V~Hole\v{s}ovi\v{c}k\'ach 2, CZ-180~00,\\ Praha 8, Czech Republic
e-mail: zasche@sirrah.troja.mff.cuni.cz \\
 \altaffilmark{2} Variable Star and Exoplanet Section of the Czech Astronomical Society, Vset\'{\i}nsk\'a
941/78,\\ CZ-757 01 Vala\v{s}sk\'e Mezi\v{r}\'{\i}\v{c}\'{\i}, Czech Republic }

%\Received{Month Day, Year}
%\end{Titlepage}

\begin{abstract}
 \noindent We present the discovery of a rare system detected in the TESS data showing three different
eclipsing-like signals. TIC 452991707 \& TIC 452991693 seem to be the second such system on the sky,
whose two components separated about 16$^{\prime\prime}$ are gravitationally bounded, or comprise a
co-moving pair. The three periods detected from the TESS data are: $P_A = 1.46155$~d, $P_B =
1.77418$~d, and $P_C = 1.03989$~d, respectively. The A and B periods belong to TIC 452991707, while the
C comes from the component TIC 452991693. The pair A shows the deepest eclipses, and its orbit is very
slightly eccentric. The third period C has lowest amplitude (eclipsing or ellipsoidal nature), but
originates from TIC 452991693, which is connected to A+B because both visual components share similar
proper motion and distance. Long-term collection of data from older photometry from various surveys
also shows that the two inner pairs A and B orbit around their barycenter. Its period is probably of a
few years, but for a final derivation of its orbital parameters one needs more up-to-date data. Hence,
we call for new observations of this amazing system.
 \end{abstract}


\section{\bf{Introduction}} \label{intro}

The classical eclipsing binaries are being used for decades for deriving the precise stellar
parameters such as radii, masses or luminosities. Also can be used as distance indicators, even
outside of our Galaxy (see e.g. Graczyk et al. 2014, or Paczynski 1997). We can also study the
stellar evolution in them, calibrating different models, detect the tidal interaction in closer
systems or discover the additional components in these systems (see e.g Guinan \& Engle 2006, and
Borkovits et al. 2016). However, some of the eclipsing binaries were found worth of studying simply
because they show some unexpected behaviour and only closer look to them would be able to reveal
what is actually going on in them.

Twenty years ago there was not known any multiple stellar system with two eclipsing binaries. Since the
discovery of V994~Her (Lee et al. 2008) proving its doubly eclipsing nature detecting the two eclipsing
periods there. Just two years ago it was not known any sextuple system with three eclipsing binaries
inside. Since the discovery of TIC~168789840 (Powell et al. 2021), and its three eclipsing periods. And
now, it seems that on the whole sky there are probably much more similar systems like this containing
three eclipsing binaries. Should we expect also a discovery of an octuple system with four eclipsing
binaries soon?
\\

\section{\bf{System discovery}}

The system TIC 452991707 \& TIC 452991693 was discovered during our scanning of the TESS photometry of
interesting sources, trying to identify there some additional eclipses, remarkable systems, triples
with coplanar orbits causing extra eclipses on the long orbit, etc. There exist a huge group of
undiscovered and still unstudied systems showing two sets of eclipses in the TESS data waiting to be
discovered and analysed. And this is one of them.

The TESS photometry was extracted using raw TESS data with the {\tt{lightkurve}} tool (Lightkurve
Collaboration et al. 2018). After then, in both sectors 17 and 18 also an additional subtraction of
longer trends was done with a polynomial fitting.

\begin{figure}
 \centering
 \includegraphics[width=0.35\textwidth]{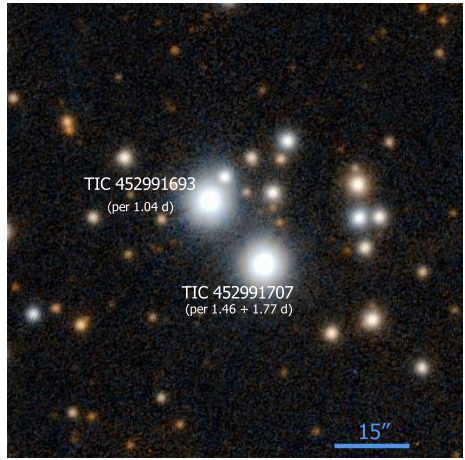}
 \caption{The image of the field and of both components of the system obtained from
 the Pan-STARRS Survey DR1 (Chambers et al. 2016).}
 \label{image}
\end{figure}

The most problematic issue when reducing the photometric data seems to be the fact that the TESS
satellite provides only poor angular resolution (pixel size is 21$^{\prime\prime}$) and the signal from
two different close-by sources (about 16$^{\prime\prime}$) fall into the same pixel. Hence, its
photometry is combined together and such blending is usually problematic when having no other
ground-based follow up observation with higher angular resolution. The field of both close components
is being plotted in Fig. \ref{image}.

\begin{figure}
 \centering
 \begin{picture}(350,400)% width and height of the picture
  \put(30,280){
  \includegraphics[width=0.49\textwidth]{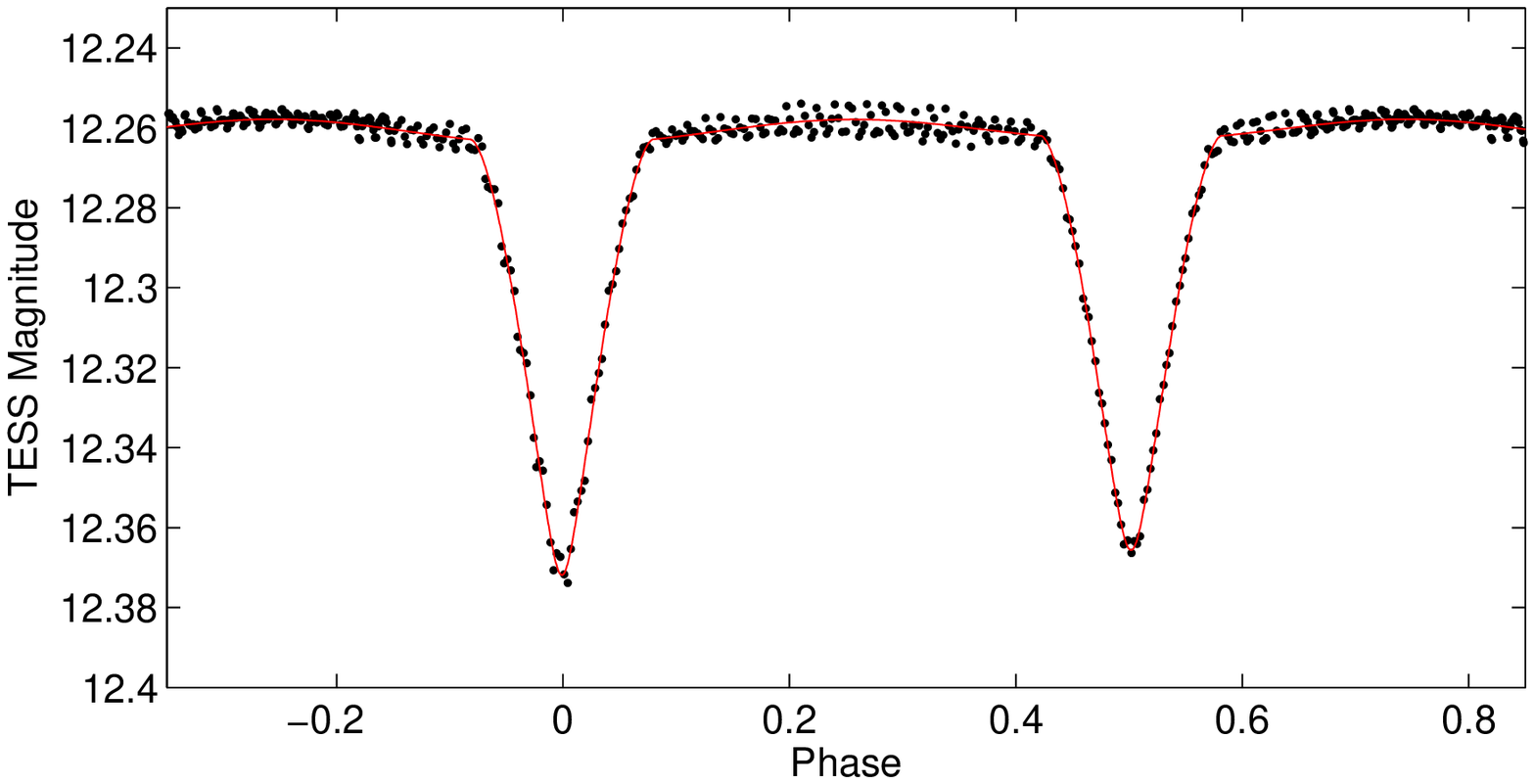}}
  \put(30,140){
  \includegraphics[width=0.49\textwidth]{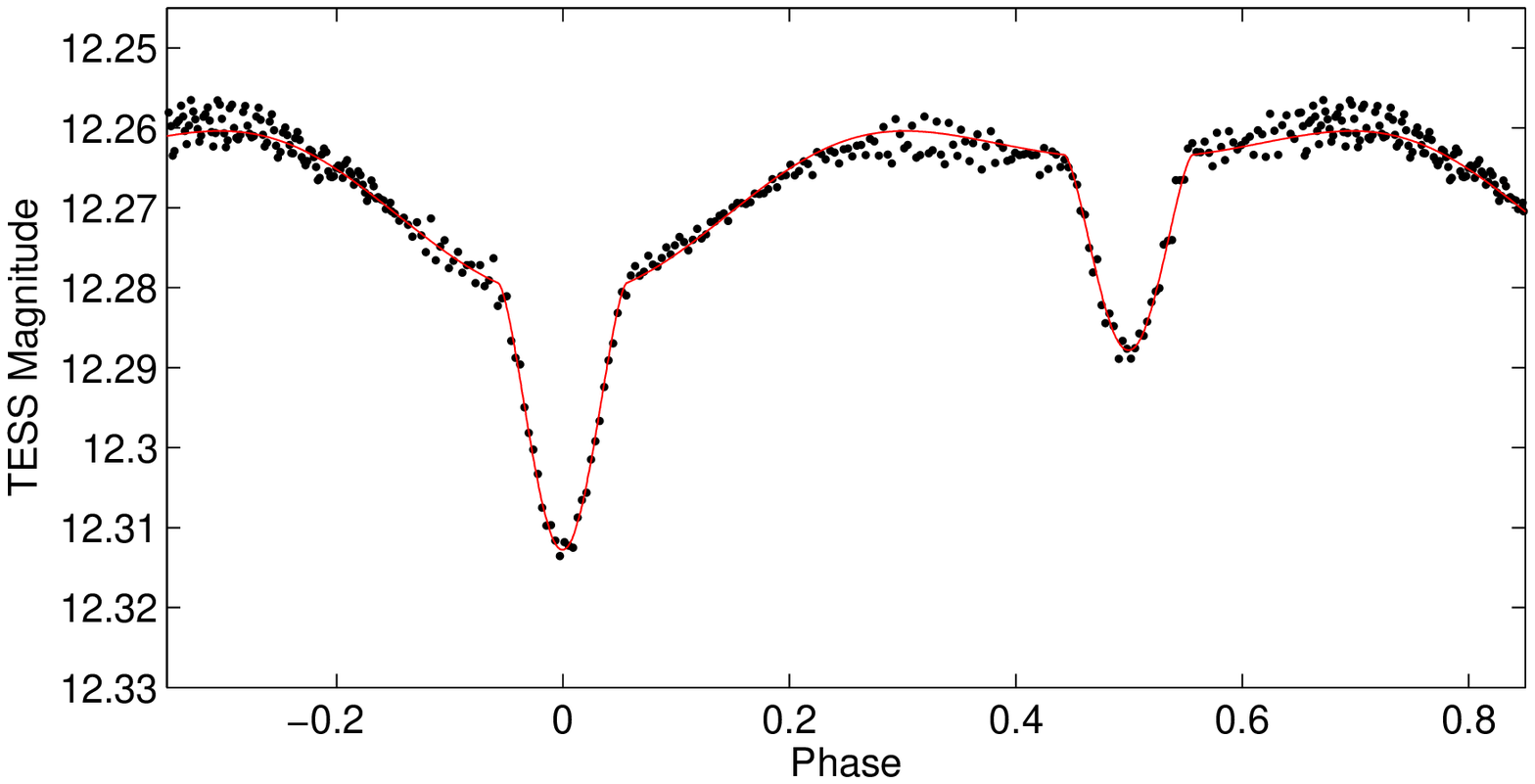}}
  \put(30,0){
  \includegraphics[width=0.49\textwidth]{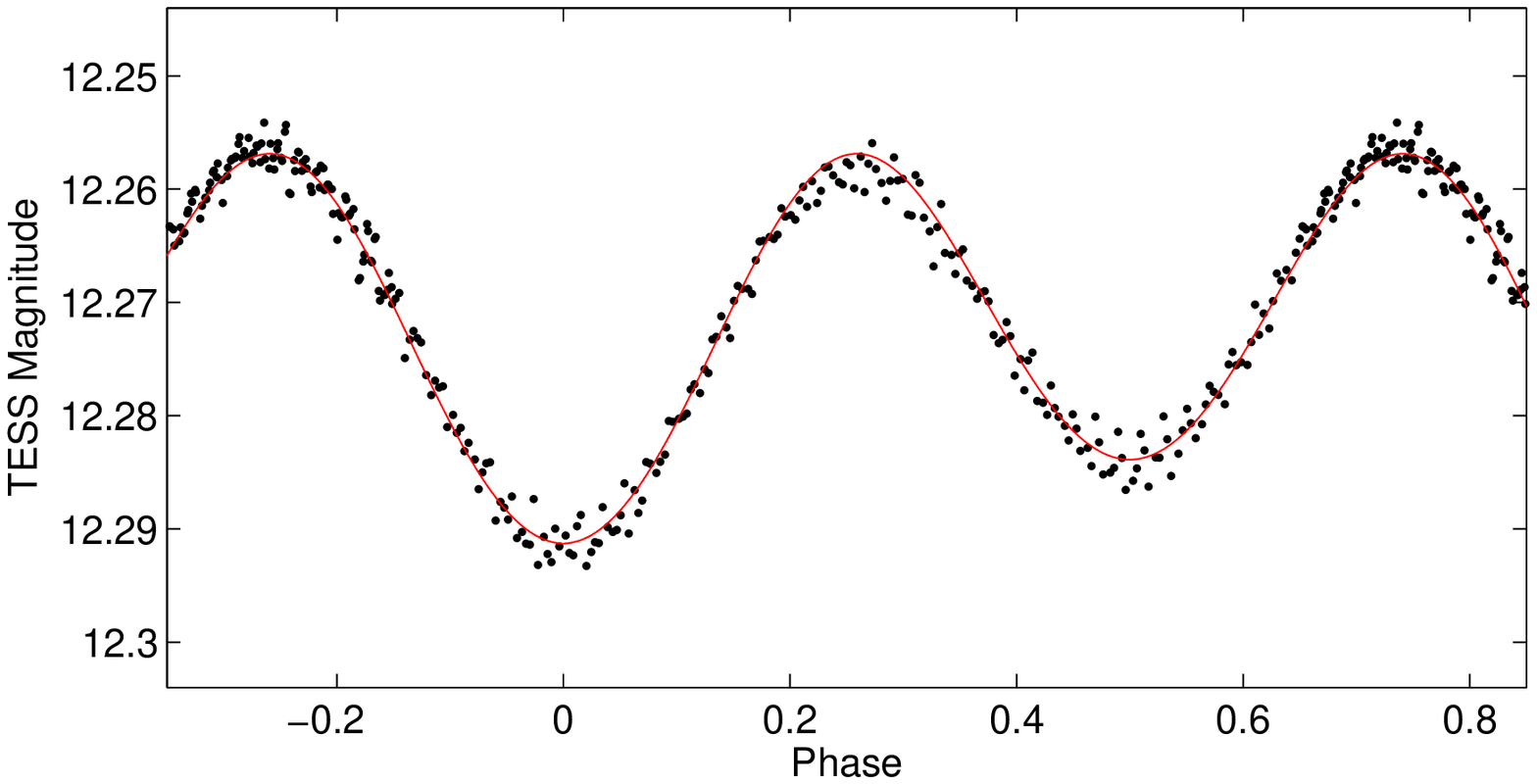}}
   \put(145,310){\small TESS: Pair A}
   \put(145,170){\small TESS: Pair B}
   \put(145,30){\small TESS: Pair C}
  \end{picture}
  \caption{Light-curve fits of all three pairs detected in the TESS data.}
  \label{FigLC}
 \end{figure}

Already from the TESS photometry there were detected two main periods and after their subtraction also
the third one. We named the most pronounced eclipse as pair A (having about 1.46 day period), while the
other one slightly shallower as pair B (period 1.77 days). The third periodicity C of about 1.04 days
is harder to detect there, due to its lowest amplitude. Consider that the pair A has the amplitude of
photometric variation over 0.1~mag, pair B over 0.05~mag, while pair C of about 0.035~mag. When
plotting the residuals after subtraction of both A and B pairs, the periodicity of pair C is clearly
visible. However, its nature would not be so clear at the first sight. One can for instance doubt
whether it is an eclipsing, or a pulsational pattern with 0.52 day period. But when analysing the data
in more detail taking into account the most plausible fit of A and B, then one can clearly see that
both primary and secondary eclipses are clearly of different depths (see below Fig. \ref{FigLC}). Hence
to conclude, such a light curve shape shows that it cannot be caused by pulsations, but rather caused
by ellipsoidal variations of a close binary.
\\

 \section{\bf{The TESS data}}

\begin{figure}
 \centering
 \includegraphics[width=0.55\textwidth]{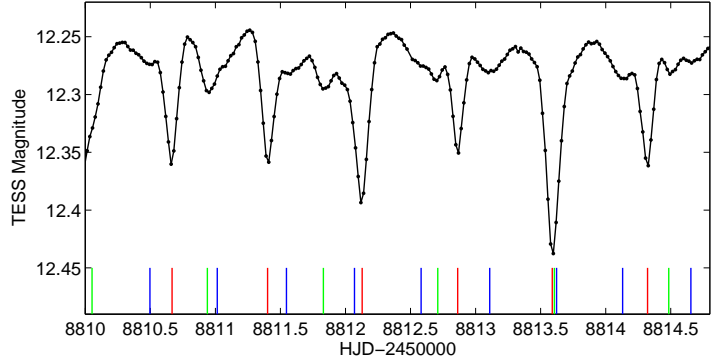}
 \caption{Sample of the light curve from TESS, sector 18. Moments of eclipses plotted in the bottom part
 of the plot: of pair A marked in red, pair B in green, while pair C in blue colour.}
 \label{TESSdata}
\end{figure}

As one can see from the illustrative picture of the light curve obtained with the TESS satellite in
Fig. \ref{TESSdata}, the combined light curve is rather complicated and the star cannot be easily
detected from the ground-based data. The TESS data provide continuous coverage of many days in the raw,
but suffers from its large pixels. We cannot easily separate the signal from the two sources,  and have
to analyse the whole light curve of all three signals together.

This task was done iteratively in several steps. In each step a preliminary fit of a particular pair
was subtracted and the residuals were analysed for the other two pairs. After then, removing another
pair, trying to find the best solution (when subtracting all three light curves, one should obtain
almost stochastic behaviour of residuals around zero value).

We used the program {\sc PHOEBE} ({Pr{\v{s}}a} \& {Zwitter} 2005) for the light curve analysis, a
freely public available software based on the Wilson-Devinney method (Wilson \& Devinney 1971). Several
simplifying assumptions have to be made due to missing spectroscopy and having only very limited
information in our hands (these were mainly the assumption of synchronous rotation, and limb-darkening
coefficients being interpolated from the tables of Van Hamme (van Hamme 1993) and using the logarithmic
law). The most serious seems the be the issue of its effective temperature (of its primary component).
Several different sources provide very different values of the $T_{eff}$, ranging from 5101~K (Gaia DR2
2018) to 7017~K (Bai et al. 2019), and one cannot easily judge which one to prefer. However, thanks to
quite moderate angular separation of both sources TIC 452991707 \& TIC 452991693 of about
16$^{\prime\prime}$, we are able to obtain different photometric information ranging over large
wavelength values for both of them. These observations coming mostly from large-scale photometric
surveys, or satellite databases, can be then used to estimate the temperatures using the spectral
energy distribution (SED) fitting. We greatly acknowledge the use of freely available web services VOSA
(Bayo et al. 2008), which was used for this purpose. As a result we roughly estimated that the star TIC
452991707 is slightly cooler than TIC 452991693. We fixed the primary temperature values for TIC
452991707 as 5300~K, while for TIC 452991693 as 5600~K for the whole light curve analysis, see below.
On the other hand, as we have tested the effect of using different primary temperature values is only
very small on the resulting Eclipse-Timing Variation (ETV) diagrams and the question of the whole
architecture of the system.

The results of our fitting are given in Table \ref{TabLCfit}. This table also provides the ephemerides
for all three pairs for future observations. Both pairs A and B seem to be the detached binaries, pair
A moreover with very slightly eccentric orbit ($e < 0.005$), but the pair C is probably a contact one.
For the pair B the components should maybe better be interchanged because the secondary is larger, as
well as more massive one. However, we prefer to leave the notation as it is due to the fact that now
the primary minimum (the deeper one) is located at 0.0 phase. One can see a slight asymmetry of the
light curves B and C, clearly evident near the quadratures. This can be caused by some photospheric
spots, or maybe by some improper reduction as an artifact. We leave it as an open question, waiting for
new better data to become available.
\\

\begin{table*}%[h!]
\caption{Derived parameters for all three pairs.}
 \label{TabLCfit}
% \scriptsize
% \tiny
  \centering \scalebox{0.9}{
\begin{tabular}{c |c|c|c}
   \hline\hline\noalign{\smallskip}
                & \multicolumn{1}{c|}{pair A}    & \multicolumn{1}{c|}{pair B} & \multicolumn{1}{c}{pair C}\\[0.5mm] \hline
 $HJD_0$ [d]          &2458800.4378 $\pm$ 0.0006 &  2458801.1816 $\pm$ 0.0014  &  2458801.6632 $\pm$ 0.0016 \\
 $P$ [d]              &1.4615495 $\pm$ 0.0000008 &  1.7741813 $\pm$ 0.0000021  &  1.0398909 $\pm$ 0.0000017 \\
  $i$ [deg]           &   87.88 $\pm$ 0.41       &       67.06 $\pm$ 0.52      &    48.39 $\pm$ 3.70       \\
 $q=\frac{M_2}{M_1}$  &   0.933 $\pm$ 0.018      &       0.943 $\pm$ 0.020     &    0.88  $\pm$ 0.02        \\
 $T_1$ [K]            &    5300 (fixed)          &       5300 (fixed)          &      5600 (fixed)         \\
 $T_2$ [K]            &    5239 $\pm$ 221        &      4539 $\pm$ 196         &    4693 $\pm$ 202         \\
 $R_1/a$              &   0.254 $\pm$ 0.006      &      0.258 $\pm$ 0.010      &   0.336  $\pm$ 0.021      \\
 $R_2/a$              &   0.228 $\pm$ 0.006      &      0.250 $\pm$ 0.011      &   0.312  $\pm$ 0.019      \\
 $L_1$ [\%]           &   12.0 $\pm$ 0.9         &      26.6 $\pm$ 0.8         &   31.6 $\pm$ 1.3          \\
 $L_2$ [\%]           &    9.3 $\pm$ 0.6         &      15.1 $\pm$ 0.7         &   12.6 $\pm$ 1.2          \\
 $L_3$ [\%]           &   78.7 $\pm$ 2.2         &      58.3 $\pm$ 2.4         &   55.8 $\pm$ 6.0          \\ \hline
 \noalign{\smallskip}\hline
\end{tabular}} \\
%  \begin{flushleft}
%  \footnotesize Note: The uncertainties of individual parameters are taken from {\sc PHOEBE} only, and are usually underestimated.\\
%  \end{flushleft}
\end{table*}

\section{\bf{Other photometry}}

Besides the TESS data, there are also other sources of available photometry, obtained in ground-based
observatories. The most extensive is the data set obtained during the ASAS-SN survey (Shappee et al.
2014, and Kochanek et al. 2017). This data provides excellent time coverage over four seasons, but at
the cost of lower quality (small aperture of the telescope). However, all three pairs are visible.

Besides that, the star was also found in the database of the SWASP project (Pollacco et al. 2006),
whose big advantage is a fact that it spreads our time span back to 2007. Besides that, also the ATLAS
survey (Heinze et al. 2018) was used.

However, the most important at this place is to mention the photometric survey named ZTF (Masci et al.
2019). The reason why particularly this one is so important is the fact that it provides the best
angular resolution of its photometry due to using larger telescope. Thanks to this fact, both close
photometric sources were observed separately. And with these ZTF data we were able to identify that the
variation of A+B comes from the southern target (i.e. TIC 452991707 = UCAC4 746-005302), while the C
pair is located with the northern star (i.e. TIC 452991693 = TYC 3666-1030-1). Due to the fact that
both A and B variations reside in the same photometric point on the sky, we still classify this target
as a doubly eclipsing system.
\\

\section{\bf{System architecture}}

Having known the sources of variability of all three periods, we should ask about the architecture of
the whole system. Two main questions arise. Is there some evidence that these two close-by stars really
constitute a bound multiple system? And secondly, can we provide some proof that also the A and B pairs
are gravitationally bound together? Surprisingly, both answers are: Yes.

Both visual components are of similar type (same colours, same photometric indices -- a kind of solar
like stars). Both share about the same proper motion. And their parallax according to GAIA is also
about the same: 0.3065 $\pm$ 0.0131~mas for TIC 452991707, while 0.2985 $\pm$ 0.0123~mas for TIC
452991693, according to GAIA DR3 (Gaia Collaboration et al. 2022). Hence, we can conclude that the
stars compose a weakly bound system with very long orbital period, or at least a co-moving pair in our
Galaxy. Having no information about the precise masses, we can only roughly calculate that the
semimajor axis (projected separation leads to about 50kAU semimajor axis) and solar like stars imply
the putative period of several Myr. In less dense stellar population (it is located in the outside part
of our Galaxy) even so weakly bound system can survive for certain period of time.

\begin{figure}
 \centering
 \includegraphics[width=0.45\textwidth]{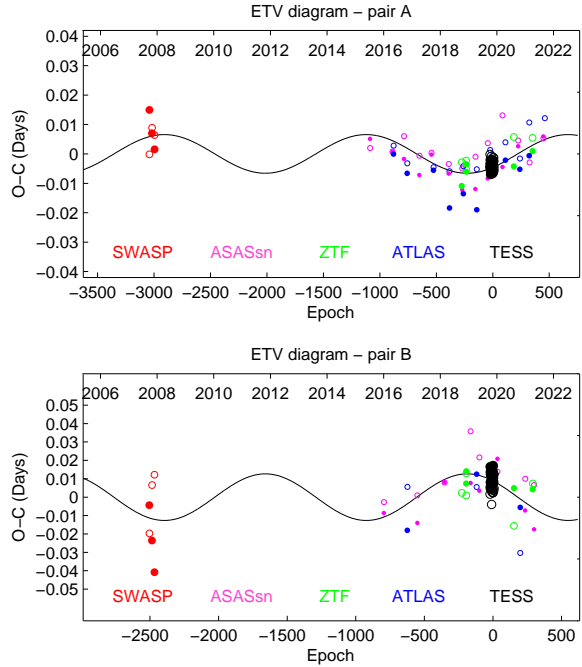}
 \caption{ETV diagram of the two inner pairs A and B with a preliminary fit of their orbit. Different colors
 denote different sources of photometry, while the filled and open circles represent the primary and secondary eclipses.}
 \label{Fig_OC}
\end{figure}

Another question of A-B orbit was resolved using all available photometry and derivation of many times
of eclipses of particular pairs in certain time epochs. A similar method was used previously for a
similar system TIC 168789840 (Powell et al. 2021), or doubly eclipsing CzeV1731 (Zasche et al. 2020).
Due to its rather shallow photometric amplitude the scatter of the pairs B and especially C are rather
poor. But still we can definitely state that the pairs A and B orbit around each other with
several-years periodicity (with the available data our analysis led to period of about 7 years).
Orbital period of the pair C remains constant. For a final derivation of orbital parameters of A-B are
the data too badly sampled in time and suffers from too large scatter. But their shape in antiphase is
clearly visible with the available data now, see Fig. \ref{Fig_OC}. Hence, its co-moving connection was
proved, which could be a sextuple configuration. At this place it would be useful to emphasize how rare
such sextuple systems are. According to our current knowledge, the most up-to-date version of the
Multiple Star Catalog MSC (Tokovinin 2018) lists only 18 sextuple and 4 proved septuple systems on the
whole sky nowadays.

We can also ask whether the two parallaxes from GAIA are reliable enough, and not biased due to the
proximity of the stars. In former Hipparcos catalogue such an effect was being discussed several times
that the close unresolved pairs yield spurious results on its parallax and proper motion. However, here
the situation is much different due to the fact that both sources are about 16$^{\prime\prime}$
distant, well beyond any such effect could possibly play a significant role. Quite a different
situation would be for the inner pair A-B, which orbits in much smaller separations. However, the two
distances derived for each of the two stars TIC 452991707 \& TIC 452991693 were derived from GAIA DR3,
which should take into account also these binary and multiple stars orbits instead of single star
solutions as in previous data releases.
\\

\section{\bf{Discussion and conclusions}}  \label{discussion}

We present a very first analysis of a sextuple system candidate with three eclipsing periods and
provide an evidence of the link between the components. Having only very limited information in our
hands and with missing spectroscopy, we still deal with a bit preliminary picture of the system.
However, the nature of the variability and its contribution to both close components is definite.

Our analysis led to finding that the mutual period of the A-B double is about $\approx$7~yrs, and we
can also derive the predicted angular separation of the A-B pair for a prospective interferometric
detection. However, thanks to its quite small parallax, this semimajor axis resulted only in about
2~mas, which is unfortunately too low for such a faint star nowadays. The spectroscopy would be also
useful for deriving the mutual orbit as well as the physical parameters of all components, however we
would need quite a lot of observing time for such a target.

One can ask, whether in such a configuration (three very close orbits of several days, an intermediate
orbit, and a long orbit of thousands of years) a system can be stable and exist for a longer period of
time. Quite recently there was published a paper about V1311~Ori (Tokovinin 2022), where its outer-most
orbit is of the order of a Myr long, and the author discussed whether it is still a bound system or a
moving group/minicluster. The structure of our system resembles e.g. the well-known sextuple system
Castor (= $\alpha$ Gem), where the respective periods also range in between a day and several thousands
of years. Many other similar hierarchies can also be found in the Multiple Star Catalog (MSC), see
Tokovinin (2018).

According to our knowledge there are about 300 doubly eclipsing systems known up to date. Only four of
them are nowadays known to be sextuples with three eclipsing binaries. For two new candidates see a
recent study by Kostov et al. (2022), while the very first sextuple system showing three eclipsing
periods was TIC~168789840 (Powell et al. 2021). Hence, a brand new field of astrophysical research is
now opening, thanks to the database of the TESS satellite.
\\

\begin{acknowledgements}
{\it Acknowledgements:} We do thank the SWASP, ZTF, ASAS-SN, TESS, and ATLAS teams for making all of
the observations easily public available. The research of P.Z. was supported by the project Progress
Q47 {\sc Physics} of the Charles University in Prague. This work has made use of data from the European
Space Agency (ESA) mission {\it Gaia} , processed by the {\it Gaia} Data Processing and Analysis
Consortium (DPAC). Funding for the DPAC has been provided by national institutions, in particular the
institutions participating in the {\it Gaia} Multilateral Agreement. This publication makes use of
VOSA, developed under the Spanish Virtual Observatory %(https://svo.cab.inta-csic.es)
project funded by MCIN/AEI/10.13039/501100011033/ through grant PID2020-112949GB-I00. VOSA has been
partially updated by using funding from the European Union's Horizon 2020 Research and Innovation
Programme, under Grant Agreement no 776403 (EXOPLANETS-A). This research made use of Lightkurve, a
Python package for TESS data analysis (Lightkurve Collaboration et al. 2018). This research has made
use of the SIMBAD and VIZIER databases, operated at CDS, Strasbourg, France and of NASA Astrophysics
Data System Bibliographic Services.
 \end{acknowledgements}

\end{document}